\documentstyle[12pt,epsfig]{article}
\vspace{1 cm}
\def\lapproxeq{\lower .7ex\hbox{$\;\stackrel{\textstyle <}{\sim}\;$}}
\def\gapproxeq{\lower .7ex\hbox{$\;\stackrel{\textstyle >}{\sim}\;$}}

\begin{document}

\titlepage
\begin{flushright} CCL-TR-95-006 \\
CERN-TH/95-105 \\
May 1995
\end{flushright}

\begin{center}
{\large{\bf The Structure Function \boldmath $F_2^{\gamma}(x,Q^2)$ at LEP2}}
\end{center}

\begin{center}
J.R. Forshaw\footnote{Talk presented at the Xth International Workshop on
Photon-Photon Collisions, `Photon 95', Sheffield, England. April 1995} \\
Rutherford Appleton Laboratory, \\ Chilton, Didcot OX11 0QX, England. \\
and \\
M.H. Seymour \\
Theory Division, CERN \\ CH-1211 Gen\`eve 23, Switzerland\\
\end{center}
\begin{abstract}
{\footnotesize{
The unique nature of the photon can be investigated in hitherto unexplored
kinematic regions at LEP2. We discuss the theoretical significance of deep
inelastic measurements and present a prescription that allows a theoretically
and experimentally sensible separation of the so-called `anomalous' and
`hadronic' components of the target photon. We perform preliminary studies
regarding the ability to reconstruct the $\gamma^* \gamma$ CM energy (and hence
$x$) and the usefulness of the easier to measure electron structure
function.
}}
\end{abstract}
\pagestyle{plain}
\begin{flushleft}
{\bf{1. The unique nature of the photon}}
\end{flushleft}
Deep inelastic scattering of electrons off hadronic targets teaches us a great
deal regarding the dynamical substructure of hadrons. If the target is a
photon then we are provided with a unique opportunity to examine the interplay
between the non-perturbative phenomena associated with hadronic bound states
and purely perturbative QCD. This is a result of the dual nature of the photon,
which can be seen to interact as a fundamental gauge boson or as a hadron.
In fig.(1) we illustrate the so-called `anomalous' and `hadronic' contributions
to the deep inelastic process (we show only the lowest order QED contribution
to the `anomalous' process).
\begin{figure}[h]
\centerline{ \epsfig{file=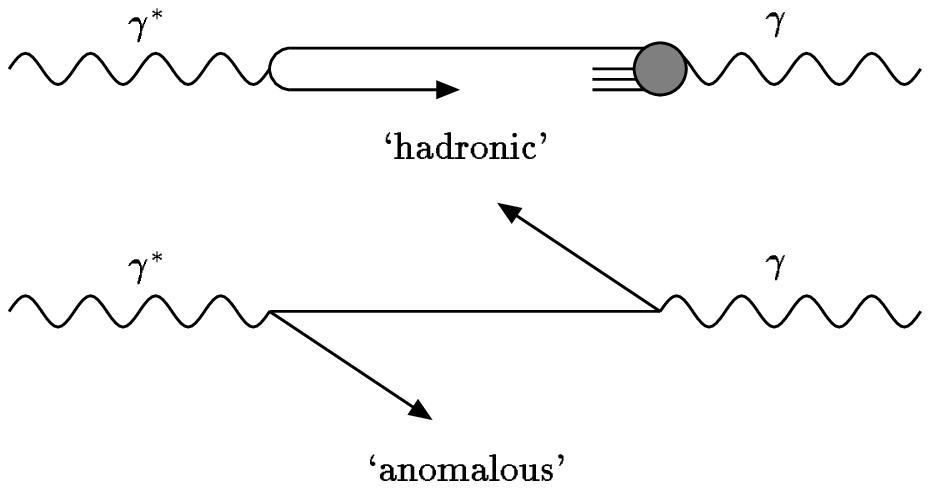,height=4.5cm}  }
\centerline{  Figure 1   }
\end{figure}
The need for a hadronic component arises, for
massless quarks, even at the QED level (i.e. $\gamma^* \gamma \to q \bar{q}$)
where one encounters a collinear divergence when integrating over the
transverse
momentum of the exchanged quark. Of course the divergence can be regulated by
introducing a quark mass, but this is perhaps unphysical since we anticipate
that the region of low transverse momentum will be subject to large
non-perturbative corrections. As a result, it is more sensible to introduce
some
factorisation scale and an associated non-perturbative component (often
modelled by vector dominance ideas). However, we expect that there will be
some deep inelastic events that are completely perturbative in nature and some
that are inherently non-perturbative. Clearly it would be interesting to
isolate such components and to study their relative contributions to the
inclusive cross section, $F_2^{\gamma}(x,Q^2)$. To quantify this separation, we
would like to propose the following definition of hadronic and anomalous events
(in fact, it is very similar to that being used already by the HERA
physicists to isolate so-called `direct' enriched events in photoproduction
\cite{xgam}).

It is appropriate to work in a frame in which the $\gamma^*$ and target
$\gamma$ are collinear. The ideal frame would be their Breit frame, in which
the virtual photon is purely longitudinal ($q_{\mu} = (0;\underline{0},Q)$),
and collides head-on with the target photon. However, since the target photon's
direction is not known exactly, we propose the Breit frame of the virtual
photon and the target electron beam. Further
studies need to be performed in order to investigate the feasibility of
performing this Lorentz transformation (which only requires sufficiently good
resolution of the tagged electron), and the extent to which using the beam
direction instead of the photon direction smears the results.
Fig.1 illustrates how a typical `hadronic'
event and a typical `anomalous' event would look in the Breit frame. The
`anomalous' events are characterised by the fact that all of the radiation is
at high $p_T$ and it is this property we exploit. Like the HERA
experimentalists, we define the variable, $x_{\gamma}$:
\begin{equation}
x_{\gamma} = \frac{ \sum (E - p_z)_{{\mathrm{jets}}} }
{ \sum (E - p_z)_{{\mathrm{all}}} }.
\end{equation}
Where the numerator sum runs over all particles in jets and the denominator sum
runs over all particles (both computed in the collinear frame  where
the +ve $z$-direction is defined by the direction of the virtual photon).
Importantly, this quantity is invariant under longitudinal boosts. For
`hadronic' events, there are no high-$p_T$ particles produced (to leading
order, the current jet and photon remnant are collinear and have zero $p_T$)
and
so the sum over jets is zero. Providing the experiments are able to identify at
least some of the target remnant, this means that $x_{\gamma} = 0$ for
`hadronic' events. For the `anomalous' events, all particles are emitted at
high $p_T$, and as such should be assigned into jets, i.e. $x_{\gamma} = 1$.
This is a remarkably clean separation of the two components and so a
considerable smearing of the distributions can be tolerated. A quantitative
definition of `hadronic' and `anomalous' is now established by making a cut on
$x_{\gamma}$.

\begin{flushleft}
{\bf{2. Small-$x$ behaviour}}
\end{flushleft}

There is already a great deal of data on the structure function,
$F_2^{\gamma}(x,Q^2)$, in the region $x \gapproxeq 0.01$ \cite{F2gam}.
LEP2 will, for the first time, allow measurement in
the lower $x$ region. In fig.2, we show the expected distribution of events in
the $x$-$Q^2$ plane given 500 pb$^{-1}$ of $e^+ e^-$ data and sensible cuts on
the tagged electron (i.e. $E_e > E_{{\mathrm{beam}}}/2$, $\theta_e >
1.7^{{\mathrm{o}}}$, where $E_e$ and $\theta_e$ are the scattered electron's
energy and angle, and $E_{{\mathrm{beam}}} = 87.5$ GeV).
\begin{figure}[ht]
\centerline{
\epsfig{file=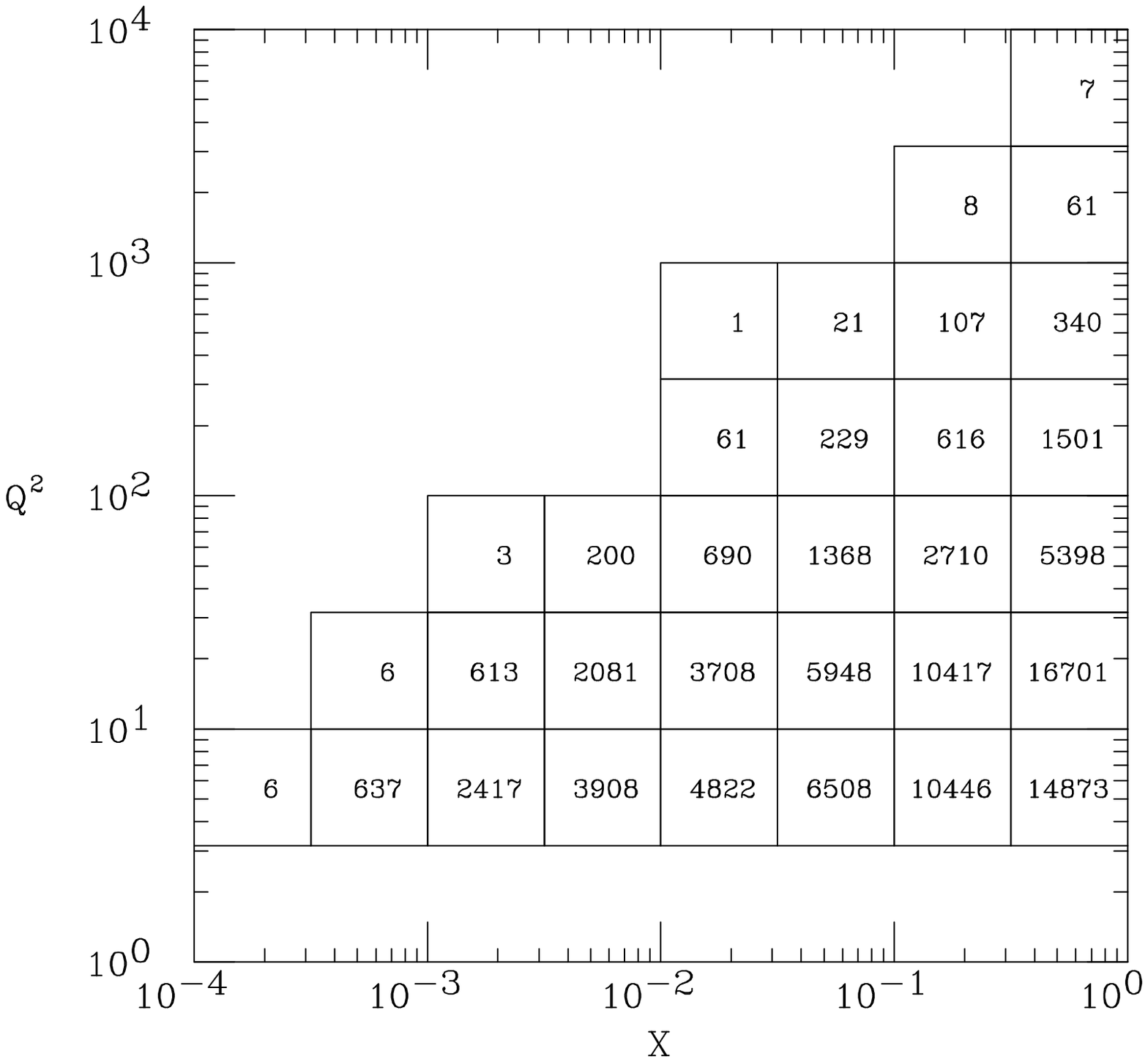,width=8.5cm,height=12cm,angle=90}  }
\centerline{  Figure 2   }
\end{figure}
As can be seen from the figure, LEP2 can expect good statistics for $x
\gapproxeq 10^{-4}$. Also notice that LEP2 is able to measure a wide
range in $x$ at a given $Q^2$. This is due to the variable target photon energy
(since it is radiated off the incoming electron).

Opening up the small $x$ domain should reveal sensitivity to the
dynamics that is responsible for the steep rise of the proton
structure function, $F_2^p(x,Q^2) \sim x^{-0.3}$.
So far this region has only been measured at HERA \cite{F2p}, so the
universality of the rise could be tested at LEP2. This rise is much stronger
than that predicted by the simple Regge pole contribution \cite{dl}
($F_2^p \sim x^{-0.08}$ at small~$x$), confirming that
perturbative physics plays an important role. Much effort has been dedicated to
identifying the nature of the large perturbative contribution. Let us outline
the basic ideas. Ultimately, the small $x$ rise is generated as a result
of many soft gluon emissions which arise due to the singular nature of the
gluon splitting function, i.e. $P_{gg} \sim 1/z$: soft gluons like to radiate
even softer gluons. The dominant contribution therefore arises from graphs like
the one in fig.3. In the conventional Dokshitzer, Gribov, Lipatov, Altarelli,
Parisi (DGLAP) approach \cite{dglap}, the splitting functions (and appropriate
coefficient functions) are expanded as power series in $\alpha_s$. At lowest
order, the solution to the DGLAP evolution equations corresponds to those
configurations where successive partons are emitted with much higher
transverse momenta than any previous emissions, so that the parent parton
of each emission can be considered collinear with the incoming hadron.
This collinear approximation leads to the summation of all
large logarithms in $Q^2$ (in leading order it is the sum of all terms $\sim
(\alpha_s \ln Q^2)^n$). However, as $x$ falls, there is increased phase space
for successively softer gluon emissions (the collinear approximation breaks
down) and terms $\sim (\alpha_s \ln z)^n$ start to become more significant.
These contributions (to the inclusive cross section) are summed up using the
formalism of Balitsky, Fadin, Kuraev and Lipatov (BFKL) \cite{bfkl}. This
latter formalism has created much interest since it provides a description of
the elusive pomeron within QCD (i.e. it is that object that determines high
energy scattering at short distances). It should however be appreciated that
the DGLAP and BFKL formalisms are not completely disjoint. By
summing an infinite subset of contributions in the expansion of the splitting
(and coefficient functions) it is possible to incorporate the leading-twist
BFKL contribution within the DGLAP approach \cite{ch,frt}.
\begin{figure}[ht]
\centerline{
\epsfig{file=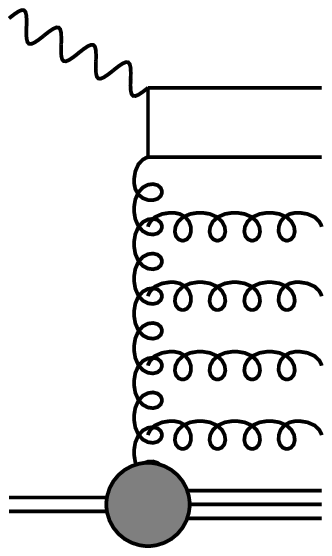,height=5.0cm}  }
\centerline{  Figure 3   }
\end{figure}

\begin{flushleft}
{\bf{3. Feasibility}}
\end{flushleft}

It is much harder to measure the $x$-dependence of the photon structure
function than that of the proton. This is because one
does not know the energy of the target photon and hence it is necessary to
reconstruct the whole of the hadronic final state in order to extract the
$\gamma^* \gamma$ invariant mass (and hence $x$). In the left-hand plot of
fig.4, we show the correlation of the observed invariant mass,
$W_{\mathrm{vis}}$, and the generated
invariant mass, $W_{\mathrm{true}}$.
\begin{figure}[ht]
\centerline{
\epsfig{file=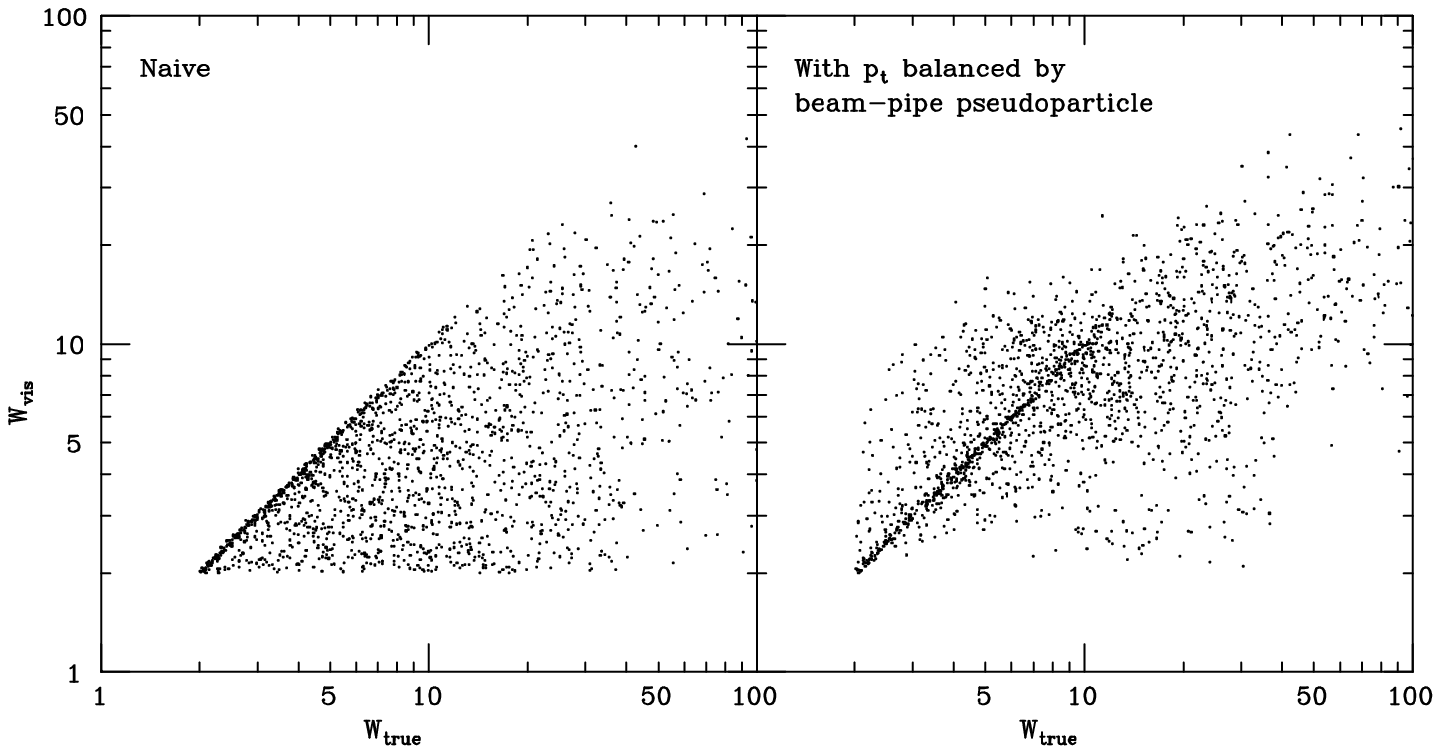,width=8.5cm,height=15cm,angle=90}  }
\centerline{  Figure 4 }
\end{figure}
We define $W_{\mathrm{vis}}$ as simply the total invariant mass of hadrons
in the region $|\cos \theta_h| < 0.97$, and show only events passing the
electron cuts given earlier, with $Q^2 < 10$ GeV$^2$ and
$W_{{\mathrm{vis}}} > 2$ GeV.
We used the HERWIG Monte Carlo event generator, but similar results have
been found using ARIADNE \cite{ll}. It is clear that the correlation is
very poor and worsens as $W_{\mathrm{true}}$ rises (i.e. $x$ falls).
At large $W_{\mathrm{true}},$ the events tend to be increasingly boosted in
the direction of the target photon, so more of the hadronic event is lost in
the beam hole, and the number of events with $W_{\mathrm{vis}} \sim
W_{\mathrm{true}}$ decreases. At the larger $x$ (lower $W_{\mathrm{true}}$)
values of the data so far collected, the correlation is good enough that a
reliable unfolding can be performed using relatively unsophisticated Monte
Carlo
programs. This will clearly not be the case at LEP2 and it is vital that effort
is devoted to establishing a more sophisticated unfolding procedure. In
the right-hand plot of fig.4, a very simple prescription (based along the lines
of an idea by John Field \cite{jf}) has been used to
define a reconstructed mass, $W_{\mathrm{recon}}$, and a vastly improved
correlation is found. The prescription utilises the information that transverse
momentum is conserved, and that the lost
mass is down the beam hole. A pseudo-particle is introduced that
carries away the missing transverse momentum (ignoring the $p_T$ of the
untagged electron) and has
longitudinal momentum just sufficient to ensure that it remains unobserved. It
is encouraging that such significant improvement is found using such a crude
algorithm.

To conclude, let us say a few words about the electron structure function,
$F_2^e(x_e,Q^2)$ (where $x_e = xz$ and $z$ is the photon energy fraction). This
measurement can be made without unfolding. In fig.5, a variety of
parametrisations for $F_2^{\gamma}(x,Q^2)$ \cite{prams} are compared along
with a similar comparison for $F_2^e(x_e,Q^2)$, both at $Q^2 = 10$ GeV$^2$.
\begin{figure}[ht]
\centerline{
\epsfig{file=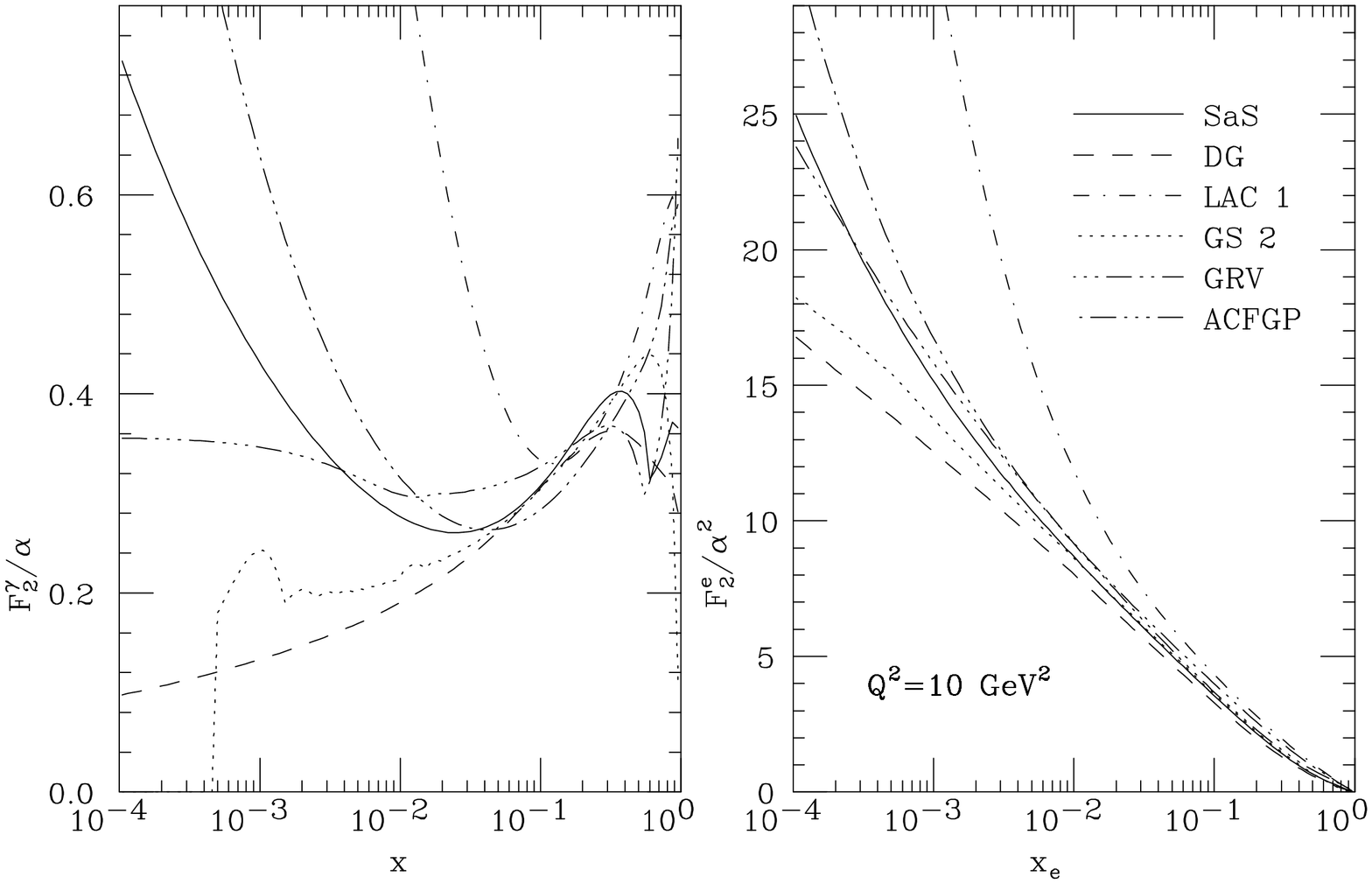,width=8.5cm,height=13cm,angle=90}}
\centerline{  Figure 5 }
\end{figure}
One can see that those parametrisations that predict very different \
behaviours for the photon at small $x$ lead to
very similar behaviours for the small $x_e$ electron structure function.
This loss of sensitivity arises because the photon flux $f_{\gamma/e}
\sim 1/z$ and so at small $x_e$ there is competition between large $z$-small
$x$ and small $z$-large $x$.
It can also be seen that not all the structure function predictions are
reliable at the $x$ and $Q^2$ values at which LEP2 will provide data (nor were
they ever intended to be so).  It is clear that this must be improved before
reliable predictions can be made for event rates or properties at LEP2 using
the full range of structure function parametrisations.

Of course another very useful measurement that avoids the need to unfold is to
measure both the electron and positron. This opens up the possibility of
measuring the virtual photon structure function, which is of significant
theoretical interest.

\end{document}